\documentclass[aps,prl,twocolumn,showpacs,superscriptaddress,nofootinbib]{revtex4}

\usepackage{graphicx}  
\usepackage{dcolumn}   
\usepackage{bm}        
\usepackage{amssymb, amsmath}   
\usepackage[usenames]{color}
\usepackage[colorlinks,urlcolor=blue,citecolor=blue,linkcolor=blue]{hyperref}

\begin{document}
\newcommand{\bra}[1]{\mbox{\ensuremath{\langle #1 \vert}}}
\newcommand{\ket}[1]{\mbox{\ensuremath{\vert #1 \rangle}}}
\newcommand{\mb}[1]{\mathbf{#1}}
\newcommand{\phipp}{\big|\phi_{\mb{p}}^{(+)}\big>}
\newcommand{\phipav}{\big|\phi_{\mb{p}}^{\p{av}}\big>}
\newcommand{\pp}[1]{\big|\psi_{p}(#1)\big>}
\newcommand{\drdy}[1]{\sqrt{-R'(#1)}}
\newcommand{\Rb}{$^{87}$Rb}
\newcommand{\K}{$^{40}$K }
\newcommand{\Li}{$^{6}$Li }
\newcommand{\LiK}{$^6$Li-$^{40}$K }
\newcommand{\na}{${^{23}}$Na}
\newcommand{\muK}{\:\mu\textrm{K}}
\newcommand{\p}[1]{\textrm{#1}}
\newcommand\T{\rule{0pt}{2.6ex}}
\newcommand\B{\rule[-1.2ex]{0pt}{0pt}}
\newcommand{\reffig}[1]{\mbox{Fig.~\ref{#1}}}
\newcommand{\refeq}[1]{\mbox{Eq.~(\ref{#1})}}
\hyphenation{Fesh-bach}
\newcommand{\previous}[1]{}
\newcommand{\note}[1]{\textcolor{red}{[\textrm{#1}]}}
\newcommand{\reddia}{\textcolor{red}{$\Diamond$}}
\newcommand{\gbox}{\textcolor{ForestGreen}{$\Box$}}
\newcommand{\bcirc}{\textcolor{blue}{$\bigcirc$}}

\title{Hydrodynamic Expansion of a Strongly Interacting Fermi-Fermi Mixture}

\author{A.\ Trenkwalder}
\affiliation{Institut f\"ur Quantenoptik und Quanteninformation,
\"Osterreichische Akademie der Wissenschaften, 6020 Innsbruck,
Austria}
\author{C.\ Kohstall}
 \affiliation{Institut f\"ur Quantenoptik und Quanteninformation,
\"Osterreichische Akademie der Wissenschaften, 6020 Innsbruck,
Austria}
\affiliation{Institut f\"ur Experimentalphysik, Universit\"at Innsbruck,
6020 Innsbruck, Austria}
\author{M.\ Zaccanti}
\affiliation{Institut f\"ur Quantenoptik und Quanteninformation,
\"Osterreichische Akademie der Wissenschaften, 6020 Innsbruck,
Austria}
\affiliation{LENS, Physics Department, University of Florence and INO-CNR, 50019 Sesto Fiorentino, Italy}
\author{D.\ Naik}
 \affiliation{Institut f\"ur Quantenoptik und Quanteninformation,
\"Osterreichische Akademie der Wissenschaften, 6020 Innsbruck,
Austria}
\author{A.\ I. Sidorov}
 \affiliation{Institut f\"ur Quantenoptik und Quanteninformation,
\"Osterreichische Akademie der Wissenschaften, 6020 Innsbruck,
Austria}
\affiliation{Centre for Atom Optics and Ultrafast Spectroscopy and ARC Centre of Excellence for Quantum-Atom Optics, Swinburne University of Technology, Melbourne, Australia}
\author{F.\ Schreck}
 \affiliation{Institut f\"ur Quantenoptik und Quanteninformation,
\"Osterreichische Akademie der Wissenschaften, 6020 Innsbruck,
Austria}
\author{R.~Grimm}
\affiliation{Institut f\"ur Quantenoptik und Quanteninformation,
\"Osterreichische Akademie der Wissenschaften, 6020 Innsbruck,
Austria}
 \affiliation{Institut f\"ur Experimentalphysik, Universit\"at Innsbruck,
6020 Innsbruck, Austria}

\date{\today}


\begin{abstract}
We report on the expansion of an ultracold Fermi-Fermi mixture of $^6$Li and $^{40}$K under conditions of strong interactions controlled via an interspecies Feshbach resonance. We study the expansion of the mixture after release from the trap and, in a narrow magnetic field range, we observe two phenomena related to hydrodynamic behavior. The common inversion of the aspect ratio is found to be accompanied by a collective effect where both species stick together and expand jointly despite of their widely different masses. Our work constitutes a major experimental step for a controlled investigation of the many-body physics of this novel strongly interacting quantum system.
\end{abstract}

\maketitle

Since the first observations of strongly interacting Fermi gases \cite{Ohara2002ooa,Bourdel2003mot} the field has produced many exciting results and provided important new insights into the many-body behavior of strongly interacting quantum matter \cite{Inguscio2006ufg,Bloch2008mbp,Giorgini2008tou}. The general interest cuts across different branches of physics, ranging from strongly correlated condensed-matter systems to neutron stars and the quark-gluon plasma.

Experimental realizations of strongly interacting Fermi gases rely on ultracold mixtures of two components with magnetically tunable $s$-wave interaction. First experiments focussed on spin mixtures of a single fermionic species with equal populations in two different Zeeman states \cite{Inguscio2006ufg}.
The introduction of population imbalance \cite{Zwierlein2006fsw,Partridge2006pap} then paved the way to the rich physics of polarized Fermi gases \cite{Radzihovsky2010ifr,Chevy2010ucp}. The recent experimental efforts to create ultracold mixtures of two different fermionic species \cite{Taglieber2008qdt, Wille2008eau, Voigt2009uhf, Spiegelhalder2009cso, Spiegelhalder2010aop, Tiecke2010bfr, Costa2010swi, Naik2011fri} have brought the field close to a new research frontier with intriguing new possibilities, e.g.\ related to novel types of superfluids and quantum phases \cite{Liu2003igs, Iskin2007sai} and to new few-body states \cite{Levinsen2009ads,Nishida2008ufg}.

In this Letter, we report on the creation of an ultracold Fermi-Fermi mixture of \Li and \K atoms, featuring the high degree of interaction control that is necessary to explore the strongly interacting regime. As a first experimental benchmark, we demonstrate the hydrodynamic expansion after release from the trap. Near the center of an interspecies Feshbach resonance, we observe two different hydrodynamic phenomena with a pronounced dependence on the interactions strength. The first one is the well-known inversion of the aspect ratio \cite{Ohara2002ooa}. The second one is a hydrodynamic drag between both species, causing their flow velocities to be equal.

We point out that both hydrodynamic phenomena find close analogies in experiments aiming at the creation of a quark-gluon plasma \cite{Braunmunzinger2007tqf,Jacak2010ctp}. Experiments of this kind study the high-energy collisions of heavy nuclei and detect the expanding collision products. In this context ``elliptic flow'' refers to an anisotropy of the expansion, which is understood as a consequence of the hydrodynamic interaction between the various collision products. The second analogy becomes evident in the transverse energy spectra of the collision products. Here it is found that heavier particles carry larger energies than the lighter ones \cite{Bearden1997cei}. Such a mass-dependence is interpreted as a result of ``collective flow'' (see e.g.\ \cite{Yagi2005qgp}), which provides another signature of the hydrodynamic nature of the expansion. The analogy between elliptic flow and the expansion of a strongly interacting Fermi gas has been pointed out already in context with early experiments on ultracold Fermi gases \cite{Ohara2002ooa,Schafer2009npf,Thomas2010tnp}. The collective flow analogy is another striking example for the fascinating relation between two fields of physics at energies differing by more than 20 orders of magnitude.

\begin{figure}
	\centering
\includegraphics[width=8.5cm]{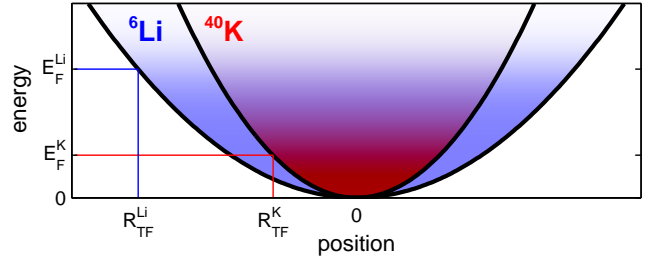}
	\caption{(color online). Illustration of the optically trapped Fermi-Fermi mixture. A small, moderately degenerate \K cloud resides in the center of the larger Fermi sea of $^6$Li. The optical trapping potential of \K is about 2.1 times deeper than the one for \Li (solid lines). The Thomas-Fermi radii $R_{\rm TF}^{\rm Li}$ and $R_{\rm TF}^{\rm K}$ of both species differ by a factor of about two.}
	\label{fig:basicsitu}
\end{figure}

The starting point of our experiments is a weakly interacting mixture of $7.5 \times 10^4$ \Li atoms and $2.0 \times 10^4$ \K atoms in an optical dipole trap \cite{trapdetails}; see illustration in Fig.\ \ref{fig:basicsitu}. The anisotropy of the trapping potential leads to a cigar-shaped sample with an aspect ratio of about 6.5. The preparation procedures are described in detail in Ref.\ \cite{Spiegelhalder2010aop}. At a temperature $T \approx 300$\,nK the Li component forms a degenerate Fermi sea with $T/T_F^{\rm Li} \approx 0.3$ and the K component is moderately degenerate with $T/T_F^{\rm K} \approx 0.7$; here the Fermi temperatures of both species are given by $T_F^{\rm Li} = 1.1\,\mu$K and $T_F^{\rm K} = 500\,$nK. The K cloud is concentrated in the center of the bigger Li cloud, with approximately equal peak densities.

Interaction control is achieved by the 155\,G interspecies Feshbach resonance, which occurs for Li in its lowest internal state ($m^{\rm Li}_f = +1/2$) and K in its third to lowest state ($m^{\rm K}_f = -5/2$) \cite{Wille2008eau,Voigt2009uhf,Naik2011fri}.
The $s$-wave scattering length $a$ can be tuned according to the standard resonance expression $a = a_{\rm bg}(1 - \Delta/(B-B_0))$ with $a_{\rm bg} = 63.0\,a_0$ ($a_0$ is Bohr's radius), $\Delta = 880\,$mG, and $B_0 = 154.707(5)\,$G \cite{Naik2011fri}.

The Li Fermi energy $E^{\rm Li}_F =k_B T^{\rm Li}_F$ represents the leading energy scale in our system. Therefore a natural condition for strong interactions of the K minority component in the degenerate Fermi sea of Li is given by $k^{\rm Li}_F|a| > 1$, where $k_F^{\rm Li} = (2 m_{\rm Li} E_F^{\rm Li})^{1/2}/\hbar = 1/(3600\,a_0)$. In terms of magnetic detuning, this condition translates to $|B-B_0| < 15\,$mG. The character of the Feshbach resonance is closed-channel dominated \cite{Chin2010fri}, but near-universal behavior can be expected throughout the strongly interacting regime \cite{Naik2011fri}.

We create the strongly interacting mixture in a transient scheme, which minimizes the time spent near resonance and thus avoids the detrimental effect of inelastic losses \cite{Naik2011fri}. We start with a weakly interacting combination of spin states with $m^{\rm Li}_f = +1/2$ and $m^{\rm K}_f = -7/2$. The magnetic field is set to the target field near $B_0$ with an estimated uncertainty as low as 3\,mG. Then we quickly convert the mixture into a strongly interacting one by flipping the spins of the K atoms to $m^{\rm K}_f = -5/2$ using a 60\,$\mu$s radio-frequency $\pi$-pulse. We immediately turn off the optical trap, releasing the sample into free space. This procedure provides well-defined initial conditions for the expansion, with the density distributions being the ones of the non-interacting system.

\begin{figure}
	\centering
\includegraphics[width=8.5cm]{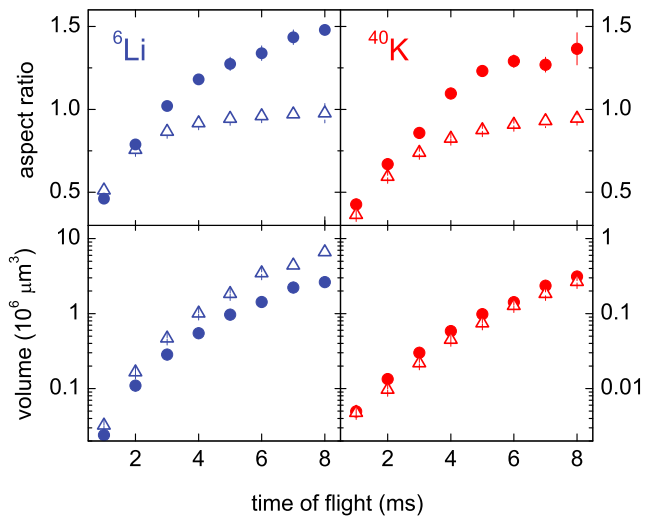}
	\caption{(color online). Expansion dynamics of the strongly interacting \LiK mixture. The upper two panels show the aspect ratios $A_{\rm Li}$ and $A_{\rm K}$, while the lower two panels display the volume parameters $V_{\rm Li}$ and $V_{\rm K}$. The closed symbols refer to the resonant case (154.709\,G), while the open symbols refer to non-interacting conditions (155.508\,G). The error bars show the statistical uncertainties of the measurements.} 
	\label{fig:expansion}
\end{figure}

In a first set of experiments, we study the expansion dynamics for a magnetic field very close to resonance ($B = 154.709$\,G). After a variable time of flight $t_{\rm TOF}$, absorption images are taken for both species and analyzed by simple two-dimensional Gaussian fits to determine their radial and axial widths, $\sigma_r$ and $\sigma_z$. In Fig.~\ref{fig:expansion} we present the resulting data in terms of the aspect ratios $A_i = \sigma^i_r/\sigma^i_z$ and volume parameters $V_i = (\sigma^i_r)^2 \sigma^i_z$, where $i = {\rm Li, K}$. For comparison, we also show corresponding measurements performed on a non-interacting sample, where the expansion proceeds ballistically and the aspect ratios asymptotically approach unity \cite{nonvsweak}.
For resonant interactions, the aspect ratios of both species, $A_{\rm Li}$ and $A_{\rm K}$, undergo an inversion, thus showing the expected hallmark of hydrodynamic behavior. Also, the volume parameters $V_{\rm Li}$ and $V_{\rm K}$ reveal striking interaction effects. While $V_{\rm Li}$ is substantially reduced by the interaction, $V_{\rm K}$ shows a small but significant increase. This observation fits to the expectation of collective flow as resulting from the hydrodynamic drag effect.

\begin{figure}
	\centering
\includegraphics[width=8.5cm]{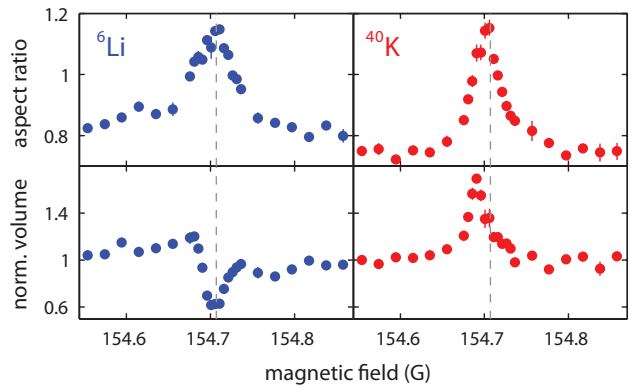}
	\caption{(color online). Magnetic-field dependence of the hydrodynamic expansion observed at $t_{\rm TOF}=4$\,ms. The upper panels show the aspect ratios $A_{\rm Li}$ and $A_{\rm K}$. The lower panels show the volume parameters $V_{\rm Li}$ and $V_{\rm K}$, normalized to their values measured for the non-interacting case. The dashed vertical lines indicate the resonance position $B_0$. The statistical uncertainties are on the order of the size of the symbols.}
	\label{fig:Bdepend}
\end{figure}

In a second set of experiments, we observe the expansion at a fixed $t_{\rm TOF} = 4$\,ms for variable interaction strength. Figure \ref{fig:Bdepend} shows the experimental data obtained for the aspect ratios and volume parameters of both species as a function of the magnetic field. Interaction effects are observed in a range with a total width of the order of 100\,mG. Deep hydrodynamic behavior, however, shows up only in a narrow range within the 30\,mG wide regime of strong interactions where $k_F^{\rm Li} |a| > 1$.
The observed magnetic-field dependence also points to interaction effects beyond elastic scattering.
In the mean field regime, the interaction is repulsive below resonance and attractive above resonance, which leads to a respective increase or decrease of the cloud size. The corresponding dispersive behavior indeed shows up in our measurements of $V_{\rm Li}$.
As another interaction effect, the maximum of the volume parameter $V_{\rm K}$ displays a shift towards lower magnetic fields with respect to both the resonance position and the maximum observed for the aspect ratio. We speculate that this shift may be related to the magnetic field dependence of the interaction energy in the strongly interacting regime.

Let us now turn our attention to another striking manifestation of the hydrodynamic drag effect, a bimodality in the spatial distribution of the expanding Li cloud. In the trap center, the Li atoms spatially overlap with the K cloud (see Fig.\ \ref{fig:basicsitu}). This inner part can together with the K atoms form a hydrodynamic core, which is surrounded by a large non-interacting cloud of excess Li atoms. In this core, multiple elastic collisions prevent the two species from separating. This leads to a slow collective expansion of the light Li atoms sticking together with the much heavier K atoms. In contrast, the expansion of the outer part of the cloud is fast and proceeds in an essentially ballistic way \cite{outercloud}.


\begin{figure}
	\centering
\includegraphics[width=8.5cm]{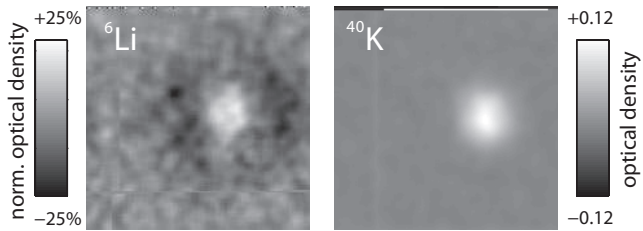}
\vspace{0mm}
	\caption{Images of the hydrodynamic core of the cloud ($t_{\rm TOF}= 4$\,ms) in the regime of resonantly tuned interactions (154.706\,G), where \Li and \K expand collectively. The image on the left-hand side is a differential image of the Li atoms, where we subtracted a reference image of the non-interacting cloud. The corresponding greyscale refers to the optical density as normalized to the maximum in the non-interacting distribution. The image on the right-hand side shows the K atoms in the strongly interacting cloud. Here the greyscale gives the optical density. The field of view of both images is $500\,\mu$m$ \times 500\,\mu$m. The images are averaged over seven individual shots.}
	\label{fig:images}
\end{figure}

Figure \ref{fig:images} shows images of the hydrodynamic core. To increase its visibility on the background of the ballistically expanding particles we show a differential Li image, where a reference image taken under non-interacting conditions is subtracted from the image of the strongly interacting cloud. This Li image is compared with a standard absorption image of the K cloud, as all K atoms are expected to contribute to the hydrodynamic core. The inner distribution detected for the Li component closely resembles the shape and size of the K cloud, supporting our interpretation of a jointly expanding Li-K cloud. The formation of this hydrodynamic core implies that particles are missing in the outer part, which undergoes a near-ballistic expansion \cite{outercloud}. Consequently, the differential Li image shows a negative signal in the outer region.

\begin{figure}
	\centering
\includegraphics[width=8.5cm]{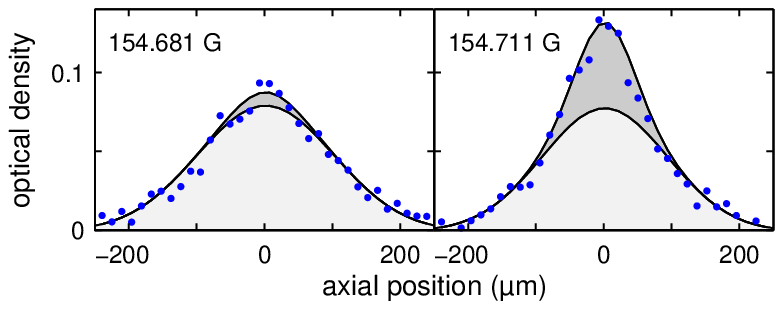}
\includegraphics[width=8.5cm]{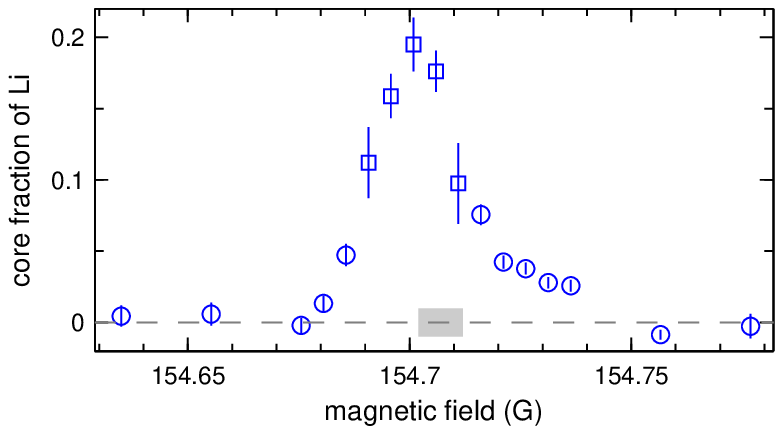}
	\caption{(color online). Bimodal distributions of \Li observed at $t_{\rm TOF} = 4$\,ms. The upper panels show two example profiles representing axial cuts through the two-dimensional distribution, obtained from a narrow strip of 15\,$\mu$m width in radial direction. The solid lines show double-Gaussian fits.
The lower panel shows the corresponding fraction of atoms in the hydrodynamic core as a function of the magnetic field. The central five points (open squares) are based on fits where the widths of the core were kept as free parameters. Further away from resonance (open circles) the bimodality is much less pronounced and such a multi-parameter fit is not applicable. Here the widths were fixed to corresponding widths of the K distribution. The error bars indicate the statistical uncertainties resulting form seven individual measurements at a given magnetic field. The grey shaded area indicates the uncertainty range for the resonance center according to Ref.~\cite{Naik2011fri}.}
	\label{fig:bimodal}
\end{figure}

We analyze the bimodal distribution of the Li cloud by two-dimensional double-Gaussian fits. Two examples for the corresponding spatial profiles are shown in the upper panels of Fig.~\ref{fig:bimodal}.
The lower panel displays the fraction of Li atoms in the hydrodynamic core that we find from such fits as a function of the magnetic field. The maximum core fraction near 20\% contains about $1.5\times10^4$ Li atoms, which corresponds to essentially all Li atoms in the overlap region.
The maximum in the core fraction of Li near 154.7\,G may be interpreted in terms of a maximum collision cross section, thus marking the exact resonance position $B_0$, but it may also point to interaction phenomena beyond elastic scattering.


The physics of interactions in the strongly interacting regime can be very rich, and more detailed investigations on the hydrodynamic core will unravel the complex many-body interactions of the system.
Besides mean-field shifts, for which we have already observed indications, strong polaronic interactions \cite{Schirotzek2009oof} or a substantial influence of pairing may be expected at sufficiently low temperatures. On the $a>0$ side, weakly bound dimers \cite{Voigt2009uhf,Spiegelhalder2010aop} may be formed either directly by radio-frequency association or indirectly by three-body recombination. We may speculate that on resonance, already at the moderate degeneracies realized in our present experiments, many-body pairs may contribute. Superfluidity may be expected at lower temperatures. All these phenomena need further detailed investigations and represent exciting future research topics.

In conclusion, we have explored a strongly interacting Fermi-Fermi mixture by studying its expansion dynamics. Our results show pronounced effects of hydrodynamic behavior, manifested in both an anisotropic expansion and in collective flow as resulting from interspecies drag. Our near-future work will be dedicated to a better understanding of the role of interaction effects, in particular to the equation of state at unitarity \cite{Gezerlis2009hlf}, and to the equilibrium and dynamics in the trap \cite{Orso2008ead}. The novel system offers many more intriguing possibilities to explore its quantum many-body physics. Already the experiments on strongly interacting spin mixtures \cite{Inguscio2006ufg,Giorgini2008tou} suggest a rich tool box of different experimental methods, such as measurements on in-situ spatial profiles, studies of collective modes, application of radio-frequency or Bragg spectroscopy, and detection of molecular condensates and fermionic pair-condensates. Moreover, the Fermi-Fermi mixture offers conceptually new possibilities through the application of species-selective optical potentials, which will allow for independent control of both components, e.g.\ for an independent manipulation of the Fermi surfaces in optical lattices \cite{Feiguin2009eps} or for the creation of mixed-dimensional fermionic systems \cite{Nishida2008ufg}.

We thank A.\ Recati, S.\ Giorgini, and S.\ Stringari for stimulating discussions and G.\ Hendl for technical assistance. We acknowledge support by the Austrian Science Fund (FWF) and the European Science Foundation (ESF) within the EuroQUAM/FerMix project and support by the FWF through the SFB FoQuS.


\begin{thebibliography}{35}
\bibitem[{tra()}]{trapdetails}
\bibinfo{note}{The trap is realized with two crossed beams derived from a
  1030\,nm single-mode laser source. The measured trap frequencies for Li (K)
  are 560\,Hz (390\,Hz) radially and 90\,Hz (57\,Hz) axially.}
\bibitem[{non()}]{nonvsweak}
\bibinfo{note}{We compared the expansion of a weakly interacting mixture ($a
  \approx a_{\rm bg}$) with the non-interacting case realized near the zero
  crossing of the scattering length ($a=0$), and found no significant
  difference.}
\bibitem[{out()}]{outercloud}
\bibinfo{note}{During the expansion some atoms in the outer cloud can interact
  with the hydrodynamic core, which also makes their distribution anisotropic.}
\end{thebibliography}


\end{document}